\documentclass[]{aa}
\usepackage{graphics}
\begin{document}
\thesaurus{4(02.16.2; 11.09.1 NGC 3521; 11.09.1 NGC 5055;
11.13.2;
11.19.2; 13.18.1)}

\title{Detection of spiral magnetic fields in two
flocculent galaxies\thanks{Partly based on observations
obtained at
 Lowell Observatory, Flagstaff, AZ (USA)} }

\author {
J. Knapik\inst{1}
 \and M. Soida\inst{1}
 \and R.-J. Dettmar\inst{2}
 \and R. Beck\inst{3}
 \and M. Urbanik\inst{1}}
\institute{ Astronomical Observatory, Jagiellonian
University, ul. Orla 171, PL30-244 Krak\'ow, Poland
\and Astronomisches Institut, Ruhr-Universit\"at-Bochum,
D44780 Bochum, Germany
\and Max-Planck-Institut f\"ur Radioastronomie, Auf dem
H\"ugel 69, D53121 Bonn, Germany}

\offprints{J. Knapik}
\mail{knapik@oa.uj.edu.pl}
\date{Received 16 June 1999/ Accepted 9 August 2000}

\titlerunning{Detection of regular magnetic fields in two
flocculent
galaxies}
\authorrunning{J. Knapik}
\maketitle

\begin{abstract}

According to the classical axisymmetric dynamo  concept,  differentially 
rotating galaxies which  lack  organized  optical  spiral  patterns  and 
density wave flows should still  have  spiral  magnetic  fields  with  a 
substantial radial component. To check this hypothesis we  observed  two 
flocculent spirals, NGC~3521 and NGC~5055, in the radio continuum (total 
power and polarization) at 10.55~GHz with a resolution of $1\farcm  13$. 
A search for traces of optical spiral patterns has  also  been  made  by 
observing them in the H$\alpha$ line and by  filtering  their  available 
blue images. 

NGC~3521 and NGC~5055 were found to possess a mean  degree  of  magnetic 
field ordering similar to that in  grand-design  spirals.  However,  the 
polarized emission fills the central region of NGC~5055 while a  minimum 
of polarized intensity was observed in the inner disk of NGC~3521.  This 
can be explained by a more uniform star formation  distribution  in  the 
centre  of  NGC~3521,  while  a  higher  concentration  of  star-forming 
activity in the nuclear region and in the rudimentary spiral ``armlets'' 
of NGC~5055 leaves broader interarm  regions  with  unperturbed  regular 
magnetic fields. Both galaxies possess regular  spiral  magnetic  fields 
with a radial component amounting to  40\%  --  60\%  of  the  azimuthal 
field. The use of beam-smoothed polarization  models  demonstrates  that 
this result cannot be produced  by  limited  resolution  and  projection 
effects. Furthermore, a large magnetic pitch angle  cannot  be  entirely 
due to the influence of rudimentary spiral-like features visible in  our 
H$\alpha$ and enhanced optical images. Thus it appears that  the  dynamo 
process is responsible for  the  radial  magnetic  field  in  flocculent 
galaxies. The measured radial magnetic field component  as  compared  to 
the azimuthal one  is  even  stronger  than  predicted  by  a  classical 
turbulent  dynamo  which  provides  arguments  in  support  for  modern, 
non-standard dynamo concepts. 

\end{abstract}

\section{Introduction}

Extensive polarization studies of nearby galaxies revealed the existence 
of spiral magnetic field configurations with a  radial  field  component 
comparable in strength to the azimuthal one (Beck  et  al.~1996).  Since 
strong shearing motions  caused  by  rapid  differential  rotation  will 
transform any  radial  structures  into  azimuthal  ones  within  a  few 
galactic  rotations,  an  effective  process  continuously  regenerating 
radial magnetic fields is  required.  The  above  findings  boosted  the 
development  of  turbulent  dynamo  theories.  According  to  these,  in 
strongly differentially rotating disks the large-scale  radial  magnetic 
field $B_r$ is generated from the azimuthal field  $B_{\phi}$  according 
to the equation (in  cylindrical  galactic  coordinates  $r,  \phi,  z$, 
Ruzmaikin et al.~1988): 

$${{\partial B_{r}\over\partial t} =
-{\partial\over\partial z}(\alpha B_{\phi}) + \beta
(\Delta\vec{B})_r } \eqno{(1)}$$

\noindent while the rotational shear converts $B_r$ back
into the azimuthal field:

$${{\partial B_{\phi}\over\partial t} = r{d\Omega\over
dr}B_r + \beta(\Delta\vec{B})_{\phi} }\eqno{(2)}$$

\noindent where $\alpha \propto <\vec{v}\cdot rot(\vec{v})>$ constitutes 
the measure of mean helicity of a small-scale turbulent  velocity  field 
$\vec{v}$ (also called  the  $\alpha$-effect),  $\Omega$  is  the  local 
angular rotation speed of the galactic disk and $\beta$ is the turbulent 
diffusion coefficient. The persistence  of  the  radial  magnetic  field 
against differential rotation constitutes a  typical  signature  of  the 
dynamo action described by eq. (1). A vertical magnetic field  component 
is generated as well, however, in realistic  dynamo  models  it  becomes 
significant at $z \ge 2-3 $~kpc (e.g. Urbanik  et  al.  1997).  This  is 
higher than the typical scale height of synchrotron emission of  $\simeq 
1$~kpc (Hummel et al. 1991), thus vertical fields are usually  not  seen 
in emission. 

Almost all nearby spiral  galaxies  known  to  possess  spiral  magnetic 
patterns (Beck et al.~1996) show strong density  waves  associated  with 
radial gas streaming motions and compression effects.  In  such  objects 
density wave flows can be at  least  partly  responsible  for  a  strong 
radial field component and high magnetic pitch angles. In some  concepts 
the non-azimuthal flows in spiral arms may even be a crucial  agent  for 
the maintenance of radial magnetic  fields  (e.g.  Otmianowska-Mazur  et 
al.~1997). Thus, the observed radial field component may result  from  a 
complex interplay of density-wave flows  and  the  $\alpha$-effect  (see 
Moss~1998), difficult to disentangle. 

\begin{table}
 \caption[]{Basic properties of NGC~3521 and NGC~5055 (from
the
 Lyon-Meudon Extragalactic Database)}
 \begin{flushleft}
 \begin{tabular}{llll}
 \hline
 & NGC~3521 & NGC~5055 & \\
 \hline
 Other names & PGC~33550 & PGC~46153, M63 &\\
 R.A.$_{1950}$ & $11^h03^m15\fs 1$ & $13^h13^m39\fs 9$ &\\
 Decl.$_{1950}$ & $0\degr 13\arcmin 58\farcs 1$ &
 $42\degr 17\arcmin 55\farcs 0$ &\\
 R.A.$_{2000}$ & $11^h05^m48\fs 9$ & $13^h15^m49\fs 2$ &\\
 Decl.$_{2000}$ & $-0\degr 02\arcmin 14\farcs 7$ &
 $42\degr 02\arcmin 05\farcs 9$ &\\
 Inclination$^{*}$ & 64$\degr$ & 55$\degr$ &\\
 Position Angle & 163$\degr$ & 105$\degr$ &\\
 Morphol. Type & SBbc & Sbc & \\
 Distance [Mpc] & 7.2$^{**}$ & 7.2$^{**}$ &\\
 Optical diameter D$_{25}$ & 10\farcm 7 & 13\farcm 0 &\\
 \hline \\
$^{*}$ $0\degr$ = face-on\\
$^{**}$ Tully (1988)
 \label{prop}
\end{tabular} \end{flushleft} \end{table}

\begin{table}
 \caption[]{Background sources removed from the radio maps
of NGC~3521}
 \begin{flushleft}
 \begin{tabular}{lll}
 \hline
 &source No.~1 & source No.~2\\
 \hline
 R.A.$_{1950}$&$11^h03^m22\fs 70$&$11^h03^m19\fs 19$\\
 Decl.$_{1950}$&$0\degr 12\arcmin 15\farcs 0$&
 $0\degr 12\arcmin 47\farcs 0$\\
 Flux density at 10.55~GHz &13.1~mJy & 13.1~mJy\\
 Degree of polarization& 18\% & 11\%\\
 Pol. position angle & 15$\degr$& 30$\degr$\\
 \hline
 \label{conf}
\end{tabular}\end{flushleft}\end{table}

Nearby spirals also have a high concentration of star formation  in  the 
spiral arms, which can act destructively on regular magnetic  fields  in 
the arms. The specific modulation  of  kinetic  helicity  and  turbulent 
diffusion by star formation  highly  concentrated  in  spiral  arms  may 
strongly influence the global magnetic field structure,  too  (Rohde  et 
al.~1998, Moss~1998). Thus, the existing data on  nearby  galaxies  does 
not provide a clear picture of dynamo-generated  magnetic  fields  in  a 
manner free from severe contamination by spiral  arms.  No  observations 
exist showing the unperturbed dynamo action in  differentially  rotating 
galaxies lacking organized spiral arms.  The  question  of  whether  the 
spiral-like magnetic fields with  a  significant  radial  component  can 
still exist in such objects (which would provide strongest argument  for 
the turbulent dynamo) has remained open up to now. In the opposite case, 
of $B_r$ being largely due to global radial flows, the magnetic field in 
the absence of organized spiral  arms  would  be  sheared  to  a  purely 
azimuthal configuration in 1 -- 2 galactic rotations. 

In this work we examine the magnetic field properties in two  flocculent 
galaxies lacking  strong  signatures  of  density  waves,  NGC~3521  and 
NGC~5055. Our aim is to check whether these  galaxies  show  any  radial 
magnetic fields (as predicted by the turbulent dynamo  concept)  in  the 
absence of systematic, ordered radial flows due  to  density  waves.  No 
existing reliable polarization data for these objects exist in  the  VLA 
archives. As the galaxies are angularly large and we are  interested  in 
the global, large-scale field geometry rather than in details  of  their 
structure, a  high  sensitivity  to  weak,  smooth,  extended  polarized 
emission was  required.  The  100\,m  Effelsberg  telescope  working  at 
10.55~GHz (to minimize Faraday effects) is a good  instrument  for  this 
purpose. 

Flocculent galaxies (including our program objects) usually possess some 
relics of a spiral structure in  the  form  of  rudimentary  ``armlets'' 
(seen e.g. in specially filtered near-infrared images, Thornley~1996) or 
spiral-like star-forming filaments and dust  lane  segments.  (see  e.g. 
optical images of NGC~3521 and NGC~5055 in Sandage~1961). NGC~5055  also 
shows some spiral-like  concentrations  of  neutral  and  molecular  gas 
(Thornley \&~Mundy 1997, Kuno et al.~1997), not accompanied  by  massive 
star formation, as usually happens in density-wave  galaxies.  All  such 
structures are substantially inclined towards the disk centre  and  thus 
have quite a large pitch angle. To  state  definitely  that  the  radial 
magnetic field component is  generated  by  the  dynamo  process  it  is 
essential to exclude the alternative possibility of  its  production  by 
local gas streaming and compression along the rudimentary arms, in local 
``miniature  density  waves''.  For  this  purpose  we  investigate  all 
possible traces of spiral structure in NGC~3521 and NGC~5055 by means of 
H$\alpha$ imaging supplemented by the analysis of available  blue  light 
images. Using beam-smoothed polarization  models  to  overcome  possible 
resolution problems (Urbanik et al. 1997) we check to which  degree  the 
magnetic field structure and in particular its radial component  can  be 
attributed to H$\alpha$-emitting, star-forming spiral-like filaments and 
dust lane segments. 

\begin{figure}
\resizebox{\hsize}{!}{\includegraphics{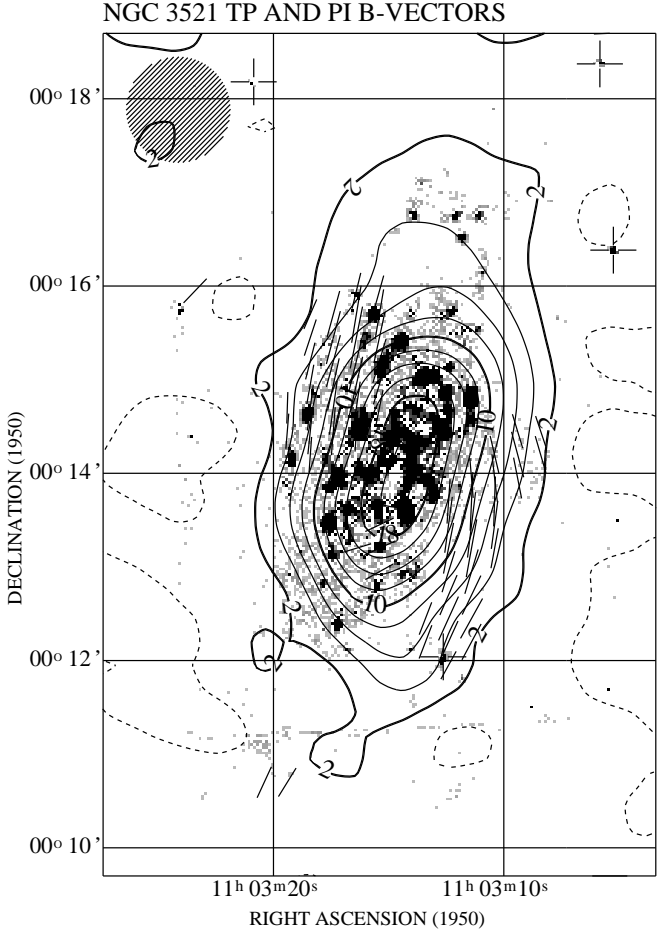}}
 \caption{
The total power contour map of NGC~3521  with  superimposed  vectors  of 
polarized intensity, both at the  original  resolution  of  68$\arcsec$, 
overlaid onto the H$\alpha$ image. The contour levels are plotted  every 
2~mJy/b.a.. The r.m.s.  noise  is  0.6~mJy/b.a.  so  the  first  contour 
corresponds to about 3.5$\sigma$ 
}
\label{n3521tp}
\end{figure}

\begin{figure}
\resizebox{\hsize}{!}{\includegraphics{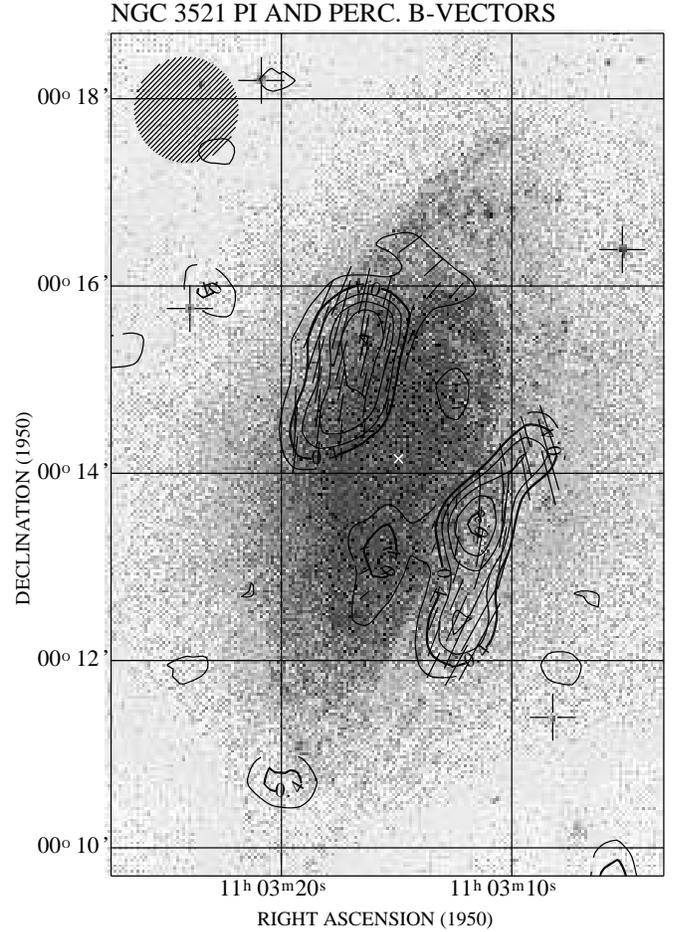}}
\caption{
The contour map of the polarized intensity of NGC~3521 with  vectors  of 
polarization degree, both at the  original  resolution  of  68$\arcsec$, 
overlaid onto the digitally enhanced blue light optical image  from  the 
``Carnegie Atlas of Galaxies'' (Sandage  \&  Bedke  1994).  The  contour 
levels are plotted from 0.3~mJy/b.a. with step 0.1~mJy/b.a.. The  r.m.s. 
noise is 0.14~mJy/b.a. so the first thick, labelled contour  corresponds 
to about 3$\sigma$. The cross denotes the galaxy's  centre  as  measured 
from our red map 
}
\label{n3521pi}
\end{figure}

\section{Observations and data reduction}

\subsection{Radio polarization}

The  total  power  and  polarization  observations  were  performed   at 
10.55~GHz using the four-horn system  in  the  secondary  focus  of  the 
Effelsberg 100\,m MPIfR telescope (Schmidt et al.  1993).  With  300~MHz 
bandwidth and 40\,K system noise temperature, the r.m.s. noise for  1\,s 
integration time and combination of all horns  is  2~mJy/b.a.  in  total 
power and 1~mJy/b.a. in polarized intensity. 

Each horn was  equipped  with  two  total  power  receivers  and  an  IF 
polarimeter resulting  in  four  data  channels  containing  the  Stokes 
parameters  I,  Q  and  U.  The  telescope  pointing  was  corrected  by 
performing cross-scans of bright point  sources  at  time  intervals  of 
about two hours. The flux calibration was  performed  using  the  highly 
polarized source 3C286. A total  power  flux  density  at  10.55~GHz  of 
4450~mJy was adopted using the formulae by  Baars  at  al.  (1977).  The 
polarized flux density was calibrated using  the  same  factors  as  for 
total power, yielding a degree of  polarization  of  12.2\%  for  3C286, 
which is in good agreement with other published values (Tabara  \&~Inoue 
1980). 

During the observations 35 coverages of NGC~3521 and 30 of NGC~5055 were 
collected in the azimuth-elevation frame. The data reduction process was 
performed using the  NOD2  data  reduction  package  (Haslam  1974).  By 
combining the information from appropriate horns, using  the  ``software 
beam switching'' technique (Morsi \&~Reich 1986) followed by restoration 
of total intensities (Emerson et al. 1979), we obtained I, Q and U  maps 
for each coverage of a given galaxy.  All  appropriate  maps  were  then 
combined  using  the   spatial-frequency   weighting   method   (Emerson 
\&~Gr\"ave 1988), followed by a digital filtering process, removing  the 
spatial frequencies corresponding to noisy structures smaller  than  the 
telescope beam. Finally the I, Q and U images  were  combined  into  the 
maps of  total  power,  polarized  intensity,  polarization  degree  and 
polarization position angles. Two polarized  background  point  sources, 
were subtracted from the  final  I,  Q  and  U  maps  of  NGC~3521  (see 
Table~\ref{conf})  prior  to  forming  the  polarization   maps.   Their 
positions and flux densities were estimated using  the  unpublished  VLA 
B-array map of NGC~3521 kindly supplied by E. Hummel.

\subsection{H$\alpha$ images}

\begin{figure*}
\resizebox{12cm}{!}{\includegraphics{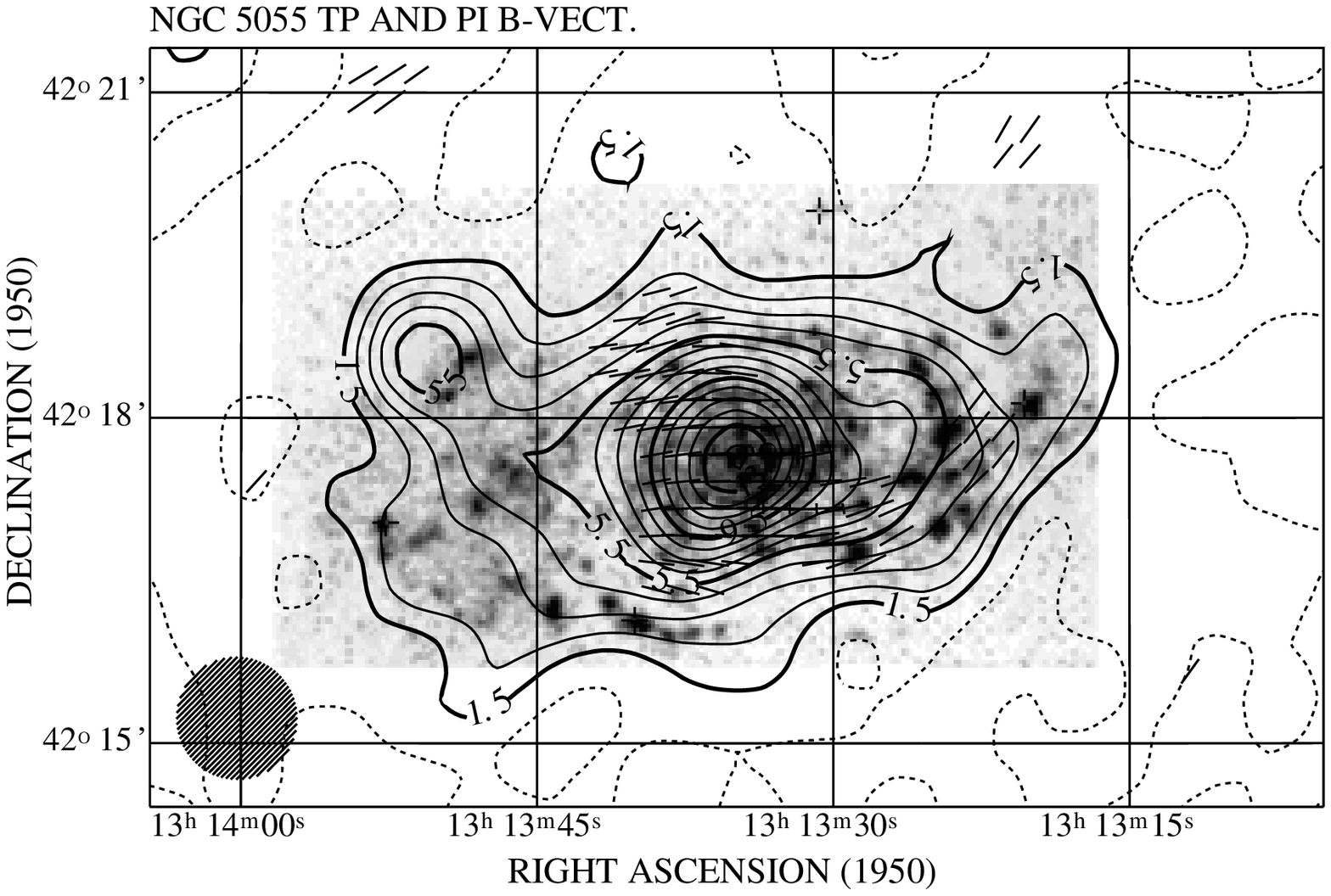}}
\parbox[b]{55mm}{
\caption
{
The total power contour map of NGC~5055  with  superimposed  vectors  of 
polarized intensity, both at the  original  resolution  of  68$\arcsec$, 
overlaid onto the H$\alpha$ image. The contour levels are  plotted  from 
1.5~mJy/b.a. with step 1.0~mJy/b.a.. The r.m.s. noise is 0.6~mJy/b.a. so 
the first contour corresponds to about 3$\sigma$ 
}
\label{n5055tp}}
\end{figure*}

\begin{figure*}
\resizebox{12cm}{!}{\includegraphics{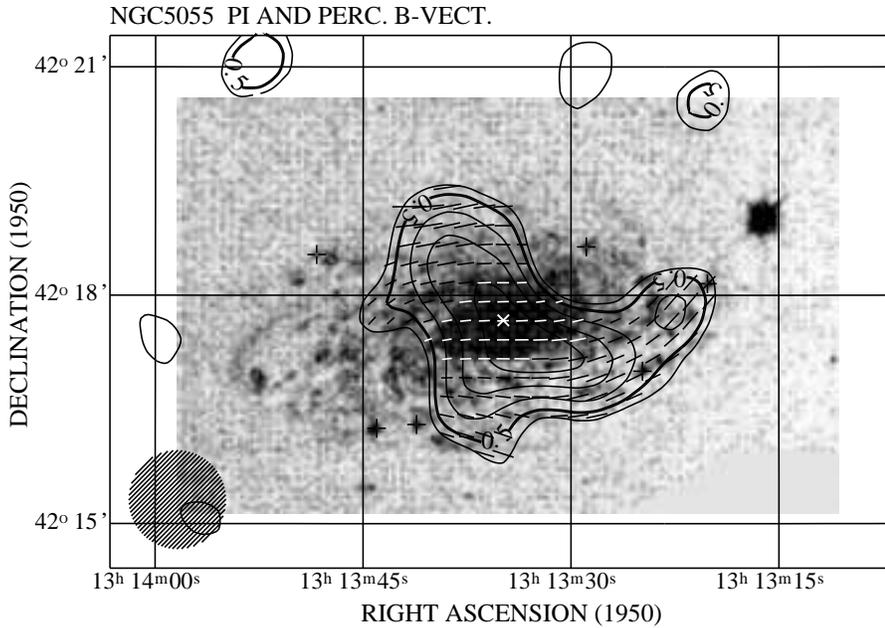}}
\parbox[b]{55mm}{
\caption
{
The contour map of the polarized intensity of NGC~5055 with  vectors  of 
polarization degree, convolved to a beam of 1\farcm 3 overlaid onto  the 
digitally enhanced optical image from the ``Carnegie Atlas of Galaxies'' 
(Sandage  \&  Bedke  1994).  The  contour  levels   are   plotted   from 
0.4~mJy/b.a. with step 0.1~mJy/b.a.. The r.m.s. noise  is  0.14~mJy/b.a. 
so the first thick, labelled contour corresponds to about 3$\sigma$. The 
cross denotes the galaxy's centre as measured from our red map 
}
\label{n5055pi} }
\end{figure*}

The optical observations were performed during four  consecutive  nights 
in  March  1996  at  the  42\arcsec\,  Hall  Telescope  of  the   Lowell 
Observatory at the Anderson Mesa site. The CCD camera used contains a TI 
800$\times$800  chip  of  15\,$\mu$m   pixels.   This   results   in   a 
field-of-view of about $4\farcm 5$ at the f/8 focus with an image  scale 
of 0\farcs 35/pixel. Therefore the objects had to  be  mapped  by  three 
overlapping fields along their major axes.  For  each  position  several 
integrations were taken in a 30\AA \ filter centered on H$\alpha$ and in 
a broad-band R. Integration times typically were 600\,sec  or  1200\,sec 
in the H$\alpha$ filter and 120\,sec or 300\,sec in R. For  the  central 
position of each galaxy a short exposure of only 20\,sec  was  taken  in 
the broad-band filter to avoid saturation of the bright nucleus. 

Since the observations were made under relatively bad seeing conditions, 
the individual images were binned by a  factor  of  two  after  standard 
reduction of the frames without loss of information. Overlapping  fields 
were combined by aligning them with reference to stars or  bright  knots 
in the overlap regions and  by  scaling  them  by  exposure  time  after 
removal  of  the  varying  night  sky  contribution.  In  combining  the 
individual frames an appropriate filter was used to avoid the  inclusion 
of saturated pixels. The continuum contribution to the intensity in  the 
H$\alpha$ filter was removed by subtracting a scaled version  of  the  R 
filter. Finally, the coordinate system for the resulting  H$\alpha$  map 
was obtained from a comparison with stars from the ST-GSC  \footnote{The 
GSC was prepared by STScI under contract with NASA (NAS5-26555)}. 

\
\section{Observational results}

The H$\alpha$  emission  of  NGC~3521  lacks  evident  signs  of  spiral 
structure. The eastern side of the disk is flanked by  a  chain  of  HII 
regions extending towards the  north  along  a  pair  of  weak  mutually 
crossing arms.  The  total  power  map  of  NGC~3521  with  superimposed 
B-vectors of polarized intensity overlaid upon  the  H$\alpha$  data  is 
presented in Fig.~\ref{n3521tp}. The galaxy shows a  bright  radio  disk 
with some extension along the peculiar dust lane in the SW disk, visible 
also in the map of Urbanik et al. (1989) and of Condon (1987). It is not 
associated with any H$\alpha$-emitting  structures  thus  it  cannot  be 
caused by local star-forming processes. The map of  polarized  intensity 
(Fig.~\ref{n3521pi}) shows two  separate  polarized  regions,  extending 
along the disk boundary and shifted clockwise from the minor  axis.  The 
observed B-vectors are roughly parallel to the faint optical spiral-like 
features and dust lanes. The SE part of the polarization map is strongly 
affected by confusing sources and should be considered with care. 

The integration of maps of the total power and  polarized  intensity  of 
NGC~3521 in elliptical rings with an  inclination  of  64$\degr$  and  a 
position angle of 163$\degr$  yields  the  10.55~GHz  total  power  flux 
density of $86\pm 12$ mJy  within  the  radius  of  $7\arcmin$  and  the 
polarized flux density  of  $2.1\pm  1.1$~mJy  within  $6\arcmin$.  This 
implies a mean polarization degree of $2.4\pm 1.3$\%. 

The H$\alpha$ map of NGC~5055 shows weak traces of spiral  structure  in 
the form of aligned chains of HII regions in the SE disk and west of the 
centre. The total power map (Fig.~\ref{n5055tp}) shows a bright  central 
region which is only barely resolved with our beam and  a  weaker  radio 
envelope. The polarized intensity in NGC~5055 (Fig.~\ref{n5055pi}) forms 
a broad lobe extending along the minor axis through the disk centre with 
a long extension to the NW. The magnetic vectors are mostly parallel  to 
the major axis. 

The integration of the total power map  of  NGC~5055  was  performed  in 
elliptical rings with an inclination of 55$\degr$ and a  position  angle 
of 105$\degr$ (see Table 1). It yields the total power flux  density  at 
10.55~GHz of $83\pm 13$ mJy within a radius of  $7\arcmin$  (similar  to 
the value obtained by Niklas  et  al.  1995),  and  the  polarized  flux 
density of $4.2\pm 1.5$~mJy inside $3\farcm  5$.  This  implies  a  mean 
polarization degree of $5.0\pm 2.5$\%. 
\
\section{Discussion}

\subsection{Global galaxy parameters}

The surface brightness  of  both  galaxies  at  10.55~GHz  implies  mean 
equipartition total magnetic field strengths of about  10.5\,$\mu$G  for 
NGC~3521  and  9.2\,$\mu$G  for  NGC~5055,  assuming  a  total   face-on 
thickness of the  radio  disk  of  2~kpc,  a  lower  energy  cutoff  for 
relativistic electrons of  300~MeV  and  a  proton  to  electron  energy 
density ratio of 100. For NGC~5055 we used a spectral index  of  $-0.79$ 
computed from our and Condon's (1987) measurements. The same  value  was 
adopted for NGC~3521, as no lower-frequency data corrected for confusing 
sources  are  so  far  available.  The  computed  field  strengths   are 
comparable to those of radio-bright nearby spirals (Beck et  al.  1996), 
thus the lack of organized spiral  arms  apparently  does  not  lead  to 
significantly weaker magnetic fields.

\begin{figure}
\resizebox{\hsize}{!}{\includegraphics{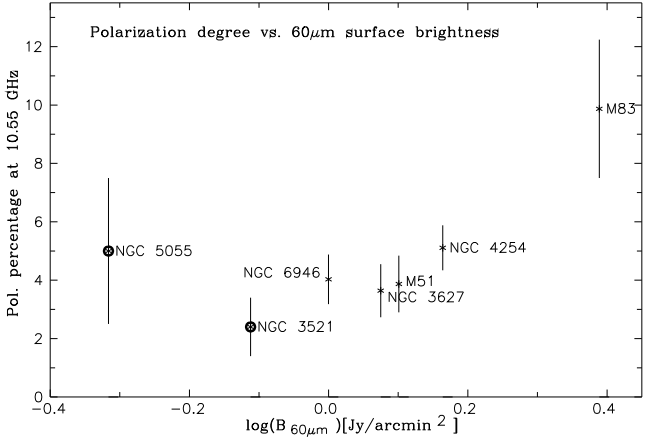}}
\caption{
A comparison of the mean polarization degree of NGC~3521,  NGC~5055  and 
nearby  grand-design  spirals  as  a  function  of  60\,$\mu$m   surface 
brightness. Before combining the U and Q data into the maps of polarized 
intensity they were convolved to the same beam relative to the  galaxy's 
optical  radius,  in  order  to  ensure  the   same   degree   of   beam 
depolarization with respect to their global magnetic  field  structures. 
The U and Q maps  are  taken  from  the  MPIfR  archives  of  Effelsberg 
polarization data. The 60\,$\mu$m  surface  brightness  is  computed  by 
dividing the IRAS flux by the face-on corrected  disk  area  within  the 
isophote of 25$^{m}/(\sq\arcsec)$ 
}
\label{corpi}
\end{figure}

Fig.~\ref{corpi} compares the mean polarization degree at  10.55~GHz  of 
NGC~3521, NGC~5055 and of grand-design spirals as a  function  of  their 
star formation efficiency (as estimated  crudely  by  the  mean  surface 
brightness at  60\,$\mu$m  taken  from  IRAS  catalogues,  Lisenfeld  et 
al.~1996). We expect that galaxies of various linear sizes  develop  the 
global magnetic field with structures closing on scales proportional  to 
the galaxy dimensions. To ensure the same degree of beam  depolarization 
with respect to galaxy-scale field structures we convolved their U and Q 
maps to the same beam size relative to the optical diameter. 

Although the polarized emission avoids regions of  high  star  formation 
inside individual galaxies (see Beck et al. 1996) there  is  no  general 
relationship between the mean  60\,$\mu$m  surface  brightness  and  the 
integrated degree of polarization (except some  deviation  exhibited  by 
M83, Fig.~\ref{corpi}). The degree of polarization of the latter  galaxy 
may be somewhat enhanced due to the magnetic field alignment  along  the 
bar (see Beck et al. 1999) or  local  gas  compressions,  seen  on  high 
contrast images as a network of sharp dust lanes filling the whole  disk 
(Malin priv. comm.). Apart from this object there is no general trend of 
IR brighter spirals being more polarized. We checked these findings  for 
a possible  bias  due  to  different  galaxy  inclinations:  no  obvious 
correlation between the polarization  degree  and  the  inclination  was 
found. 

NGC~5055 has a polarization degree higher than some grand-design spirals 
and nearly the same as NGC~4254 (see Soida et al. 1996),  which  belongs 
to the strongest polarized spirals after  M83.  NGC~3521  is  the  least 
polarized object in our sample, however its polarization degree is  only 
a factor of 0.7 lower than that of the  strong  density-wave,  similarly 
inclined object NGC~3627 (Soida et al. 1999). Taking  into  account  the 
errors of polarization degree, the flocculent galaxies studied  in  this 
work do not seem to possess a systematically lower  degree  of  magnetic 
field ordering than  that  in  grand-design  galaxies,  which  could  be 
attributed to a lack of organized spiral arms.  {\it  A  strong  density 
wave action is apparently not an important  agent  in  producing  strong 
regular magnetic fields}. 

\begin{figure}
\resizebox{\hsize}{!}{\includegraphics{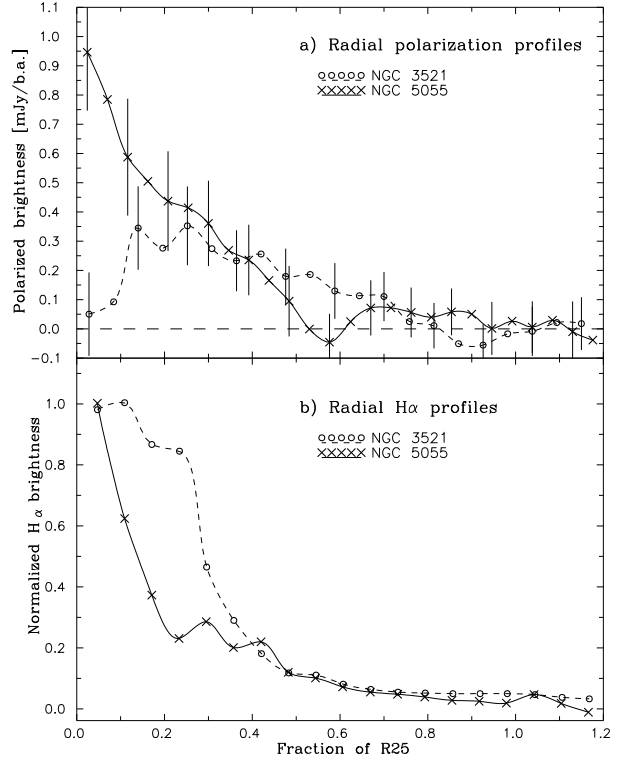}}
\caption{
a) Radial distributions of polarized intensity of NGC 3521 and NGC  5055 
integrated in rings with inclinations and major axis orientations as  in 
Table 1. The same rings $20\arcsec$ wide  are  used  for  both  galaxies 
since for the radio data the degree of data smearing  is  determined  by 
the beamsize.  b)  Radial  distributions  of  the  H$\alpha$  brightness 
integrated in rings with  orientation  parameters  from  Table  1.  Each 
H$\alpha$ profile is normalized to its maximum value. To ensure the same 
degree of data smearing relative to the optical diameter, in  this  case 
we used a ring width of $20\arcsec$ for NGC~3521 and  $24\farcs  1$  for 
NGC~5055. In both graphs the galactocentric radius is normalized to  the 
optical disk radius at the level of 25$^{m}/(\sq\arcsec)$ taken from the 
LEDA database (see Tab. 1) 
}
\label{pirad}
\end{figure}

Fig.~\ref{pirad}  compares  the  radial   distributions   of   polarized 
intensity and  H$\alpha$  brightness  for  the  studied  galaxies.  Both 
objects have (to the limits of errors) a  similar  polarized  brightness 
beyond 0.2 of the optical radius R$_{25}$. However, while the  polarized 
brightness continues to rise towards the centre in  NGC~5055,  it  drops 
suddenly  in  the  inner  disk  of  NGC~3521.  This  is  accompanied  by 
differences in the distribution of H$\alpha$ brightness.  The  H$\alpha$ 
emission in NGC~5055 is highly concentrated in the nuclear region  while 
in NGC~3521 it forms a broad ``plateau'' extending up to  0.3  R$_{25}$. 
Moreover, while star-forming processes traced by the H$\alpha$  emission 
fill  the  inner  disk  of  NGC~3521  more  uniformly,  they  are   more 
concentrated in the arm-like chains of HII  regions  in  NGC~5055.  This 
implies more efficient destruction of regular fields in the  inner  disk 
of NGC~3521 by supernova explosions, stellar  winds  etc.,  which  gives 
rise to a central depression in the polarized emission. We suspect  that 
the polarized surface brightness of the galactic disk may depend more on 
the star  formation  {\it  distribution}  rather  than  on  the  overall 
star-forming activity.  Even  rapidly  star-forming  galaxies  may  show 
highly polarized emission if star formation is strongly concentrated  in 
spiral arms, leaving broad, quiet interarm regions with  highly  ordered 
magnetic fields. 

\subsection{Structure of large-scale magnetic fields}

\begin{figure}
\vspace{3.5cm}
\resizebox{\hsize}{!}{\includegraphics{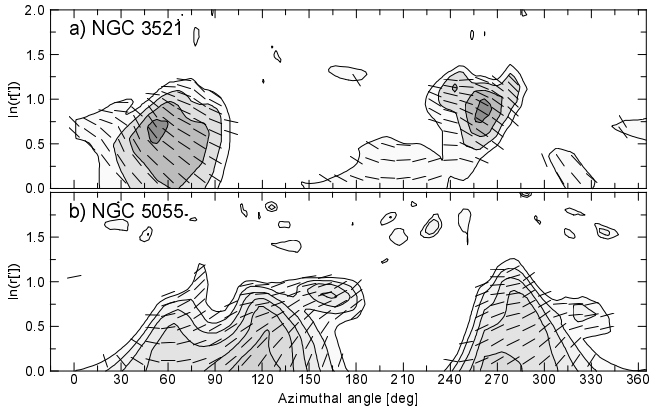}}
\caption{
Plot of the distribution of polarized intensity (contours and greyscale) 
and of the orientations of  polarization  B-vectors  in  the  azimuth  - 
$\ln(r)$ frame ($r$ is the galactocentric radius in arcmin) for NGC~3521 
(a) and NGC~5055 (b). The azimuthal angle runs  anticlockwise  from  the 
northern part of the major axis in NGC~3521 and from the western part of 
the major axis in NGC~5055 
}
\label{logpi}
\end{figure}

\begin{figure}
\vspace{3.5cm}
\resizebox{\hsize}{!}{\includegraphics{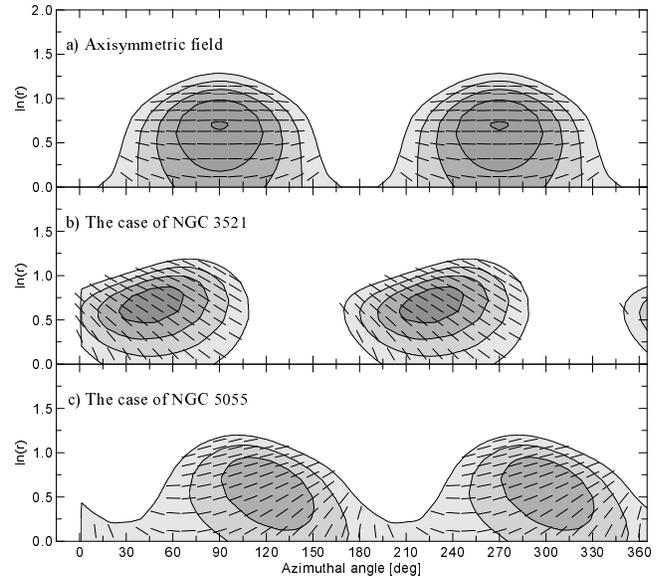}}
\caption{
Plot of the distribution of polarized intensity (contours and greyscale) 
and of the orientations of polarization  B-vectors  in  the  azimuth  -- 
$\ln(r)$  frame  ($r$  is  the  galactocentric  radius  in  arcmin)  for 
axisymmetric model containing only azimuthal magnetic field (a), for the 
dynamo-type magnetic field with regular component in the central  region 
suppressed to reproduce the polarization minimum in  NGC~3521  (b),  and 
for the case of NGC~5055 with dynamo-type regular  fields  extending  to 
the disk centre (c). In cases (b) and  (c)  we  used  a  combination  of 
poloidal and toroidal  magnetic  fields  resulting  from  the  numerical 
dynamo simulations by D. Elstner showing a  strong  vertical  component, 
with the ratio of poloidal to toroidal  fields  adjusted  to  yield  the 
pitch angles of B-vectors similar to those in real galaxies. The  models 
are convolved to a beam of 1\farcm 13 
}
\label{modpi}
\end{figure}

Fig.~\ref{logpi} shows the distribution of polarized intensity  and  the 
orientation of B-vectors as a function of azimuthal angle  in  the  disk 
plane and the natural logarithm of the galactocentric  radius.  In  such 
coordinates a logarithmic spiral appears as  a  straight  line  with  an 
inclination equal to its pitch angle. As  such  graphs  are  subject  to 
strong beam smearing effects in  central  regions,  Fig.~\ref{logpi}  is 
calculated for galactocentric radii $r  \ge  1\arcmin$.  The  region  of 
NGC~3521 between the azimuthal angles of 180$\degr$ and  240$\degr$  and 
$\ln(r) < 0.5$ is confused by the background source and we  excluded  it 
from considerations. 

\begin{figure}
\resizebox{\hsize}{!}{\includegraphics{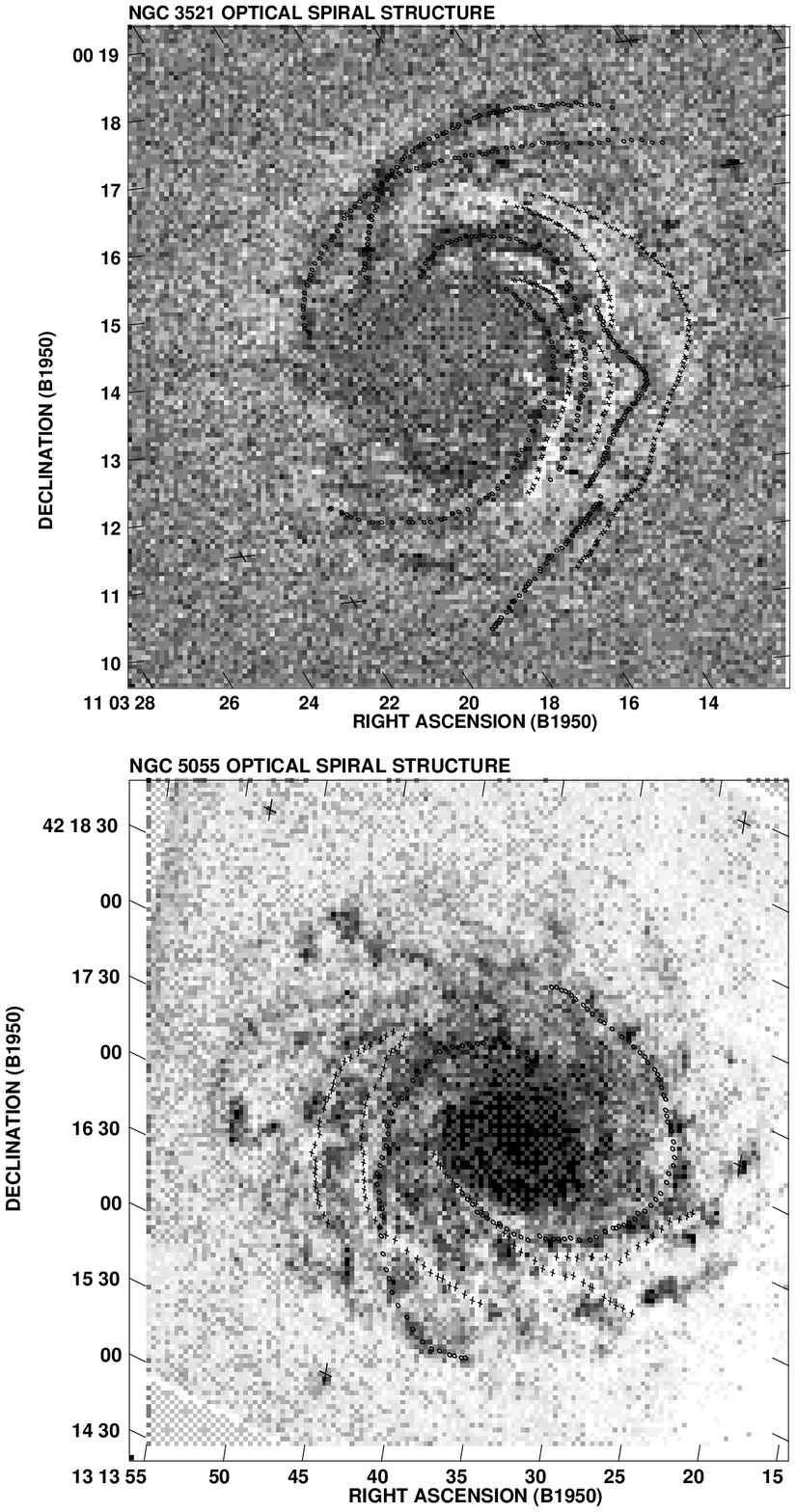}}
\caption{
The face-on deprojected, digitally filtered and  enhanced  blue  optical 
images of NGC~3521 (upper panel) and  NGC~5055  (lower  panel),  showing 
faint spiral structures. The most prominent structures  were  marked  by 
symbols -- optically bright ``armlets'' by circles and dust features  by 
crosses 
}
\label{arms}
\end{figure}


\begin{figure}
\resizebox{\hsize}{!}{\includegraphics{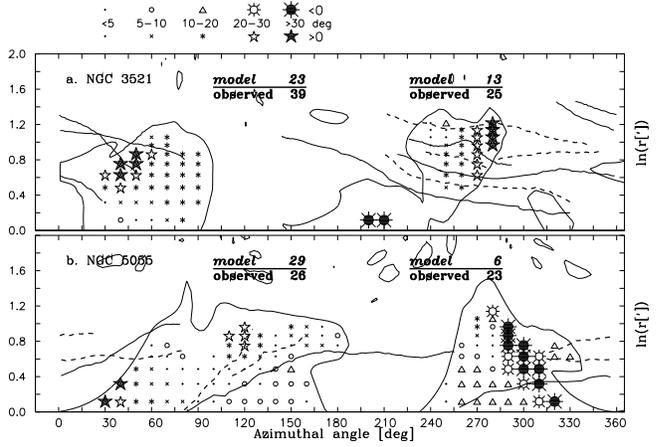}}
\caption{
The azimuth - $\ln(r)$ plot of  differences  between  observed  magnetic 
pitch angles in NGC~3521 (a) and NGC~5055 (b), and  those  obtained  for 
models assuming a magnetic field aligned with optically  bright  armlets 
and dust lanes, as marked in Fig.  12  but  having  the  same  polarized 
intensity distribution as observed. Various  symbols  denote  particular 
intervals of these differences (see the Figure legend). Pitch angles are 
computed within the second contour given in Fig. 8, for  comparison  the 
outline of the first contour is plotted as well. The dust lanes  (dashed 
lines) and optically bright ``armlets'' (solid lines)  are  also  shown. 
The error of pitch angles  is  about  5$\degr$  in  the  inner  part  of 
polarized lobes and about 7$\degr$ in their  outer  regions,  rising  to 
about 9$\degr$ in polarized extensions stretching to the right  of  main 
lobes in NGC~5055. The numbers near the top of each panel show the mean 
pitch angles: modeled  (top  line)  and  observed,  both  averaged  over 
particular polarized lobes 
}
\label{lgar}
\end{figure}

In both galaxies the magnetic pitch angles $\psi$  differ  significantly 
from zero,  their  mean  absolute  values  are  $31\degr\pm  4\degr$  in 
NGC~3521 and $24\degr\pm  4\degr$  in  NGC~5055,  thus  on  average  the 
observed radial magnetic field component constitutes  respectively  60\% 
and 42\% of the azimuthal component. In  specific  polarized  lobes  the 
mean pitch angles are: in NGC~3521 $39\degr\pm 5\degr$ in  the  NE  lobe 
and  $25\degr\pm  5\degr$  in  the  SW  one,  in  NGC~5055  we  obtained 
$23\degr\pm 5\degr$ in the northern and NE disk and $26\degr\pm  5\degr$ 
in southern and SW disk. In each galaxy the  polarized  intensity  peaks 
are placed symmetrically on both  sides  of  the  disk  but  shifted  in 
azimuth from the minor axis according to the sign of the pitch angle  of 
magnetic vectors. The azimuthal angle of the shift  is  similar  to  the 
mean pitch angle. This shift constitutes a geometrical effect caused  by 
a non-zero radial magnetic field (see Urbanik et al. 1997) and  provides 
further evidence for the existence of a radial field component. 

We checked the above  findings  for  possible  effects  of  our  limited 
resolution and the influence of vertical magnetic fields projected  onto 
the sky plane in rather highly inclined objects.  For  this  purpose  we 
used  beam-smoothed  polarization  models  based  on   three-dimensional 
dynamo-generated fields with a strong vertical component, selected  from 
a large library of dynamo results prepared by D. Elstner  in  course  of 
modeling works described by Urbanik et al. (1997). A  number  of  models 
were computed with various intrinsic magnetic  pitch  angles.  Obtaining 
the dynamo-generated magnetic fields with  a  given  (especially  large) 
pitch  angle  by  solving  the  dynamo  equations  numerically  requires 
extremely elaborate calculations. Our aim was only to check for possible 
biases introduced by  resolution  and  projection  effects  rather  than 
fitting concrete physical  dynamo  models  to  our  data,  thus  it  was 
sufficient to regulate the intrinsic magnetic pitch angle by multiplying 
the toroidal field strength in the chosen model by a constant factor.  A 
reference model containing pure azimuthal field was  computed  as  well. 
The models of polarized emission smoothed to the beam of 1\farcm 13 were 
obtained by integrating Stokes I, Q and U  parameters  as  described  by 
Urbanik et al. (1997). Separate models were computed  for  NGC~3521  and 
NGC~5055, taking into account their inclinations from  Tab.  1  and  the 
fact that the  central  region  of  NGC~3521  is  unpolarized  while  in 
NGC~5055 the polarized emission fills the whole inner disk.  The  radial 
exponential  scale  length  of  CR  electron  density  was  adjusted  to 
reproduce  the  radial  distribution  of  polarized  intensity.  In  the 
$z$-direction a Gaussian relativistic electron density distribution with 
a scale height of 1~kpc was adopted. The  simulated  maps  of  polarized 
intensity were analyzed in the azimuth -- $\ln(r)$ frame in the same way 
as those of observed galaxies. 

Results displayed in Fig.~\ref{modpi}a demonstrate that purely azimuthal 
fields can generate some inclined B-vectors. However, their pitch angles 
are of opposite sign and of the same absolute value  on  both  sides  of 
each lobe. The B-vectors are perfectly azimuthal  inside  the  polarized 
lobes, which are always  peaking  at  the  minor  axis.  This  behaviour 
results from the geometry of the azimuthal field,  whose  projection  is 
always symmetrical with respect to the minor axis, with the  maximum  of 
the regular field perpendicular  to  the  line  of  sight  ($B_{\perp}$) 
falling exactly at the minor axis. A non-zero pitch angle  of  the  same 
sign throughout the whole polarized lobes can be obtained  only  when  a 
poloidal (thus also radial) magnetic field is allowed (Fig.~\ref{modpi}b 
and c). Only in this case it was possible to obtain a shift of polarized 
intensity maxima from the minor axis, resulting  from  the  geometry  of 
magnetic lines. If a radial magnetic field component  is  added,  making 
the magnetic lines deviate from the azimuthal direction by a given pitch 
angle $\psi$, the maximum $B_{\perp} $ occurs at an  azimuthal  distance 
from the minor  axis  roughly  equal  to  $\psi$,  as  can  be  seen  in 
Fig.~\ref{modpi}b and c (see also Urbanik et al. 1997). 

If the regular magnetic field in the central region  is  suppressed,  no 
radially  directed  B-vectors  are  obtained  (Fig.~\ref{modpi}b).  If 
regular fields extend to the centre,  the  groups  of  radial  B-vectors 
occur close to azimuthal angles of $0\degr$ and $180\degr$, thus running 
across the centre along the major axis. This picture  agrees  well  with 
Fig.~\ref{logpi}a (no radial fields observed in NGC~3521) and b  (radial 
magnetic fields in  NGC~5055  especially  obvious  at  azimuthal  angles 
$0\degr$ -- $30\degr$). In the latter case our model reproduces well the 
characteristic pitch angle reversal in the mentioned range of  azimuthal 
angles. There is thus no  need  to  invoke  any  bisymmetric  fields  to 
explain the B-vectors crossing the central region in NGC~5055. 

Table~\ref{pitc} compares the mean intrinsic magnetic pitch  angles  for 
the whole series of adopted model magnetic fields  with  those  measured 
from a  beam-smoothed  polarization  model.  As  in  dynamo  models  the 
intrinsic pitch angle changes with the distance from the disk plane $z$, 
the observed values  of  $\psi$  depend  on  the  vertical  CR  electron 
distribution. To account for this effect and to single  out  effects  of 
projection and resolution the  intrinsic  pitch  angles  were  averaged, 
weighting them with the assumed vertical  distribution  of  CR  electron 
density. The observed pitch angles of B-vectors were averaged over  each 
polarized lobe in the Stokes QU plane (as the radio telescope does). 

For observed pitch angles smaller than $40\degr$ (as found  in  NGC~3521 
and NGC~5055) there is almost  perfect  agreement  between  assumed  and 
measured values of $\psi$. For larger pitch  angles,  the  beam-smoothed 
values become higher than the assumed  values  by  a  few  degrees.  The 
differences  become  large  for  observed  pitch  angles  greater   than 
$60\degr$. The effect is due to  an  increasing  role  of  the  poloidal 
magnetic field with respect to the azimuthal field which also  causes  a 
greater contribution of the sky-projected, beam-smoothed vertical  field 
component. It follows from Tab.~\ref{pitc} that the  intrinsic  magnetic 
pitch angle in both galaxies cannot be significantly  smaller  than  the 
observed values. In particular, the resolution  and  projection  effects 
are unable to produce the observed radial fields out of purely azimuthal 
ones.

\begin{table}
\caption[]{A comparison of intrinsic magnetic pitch angles
to those
measured from beam-smoothed polarized intensity maps}
 
\begin{center}
\begin{tabular}{rrr}
\hline
model&intrinsic& measured\\
&$\psi [\degr]$& $\psi [\degr]$\\
\hline
1 & 0.3 & 0.2 \\
2 & 2.7 & 2.4 \\
3 & 12.7 & 12.1 \\
4 & 22.3 & 23.1 \\
5 & 33.8 & 40.5 \\
6 & 45.6 & 64.9 \\
7 & 50.0 & 77.8 \\
\hline
 \label{pitc}
\end{tabular}
\end{center}
\end{table}

\subsection{Traces of  spiral  structure  and  the  origin  of  magnetic 
fields} 

Using our H$\alpha$ maps and the digitally enhanced blue optical  images 
(Fig.~\ref{arms}) of NGC~3521 and NGC~5055 we are  able  to  trace  weak 
signs of spiral structure. Though both galaxies exhibit  similar  traces 
of a spiral pattern in the near-infrared (Thornley 1996), NGC~3521 shows 
only weak signs of spiral structure in its H$\alpha$ emission. The inner 
disk is filled uniformly with chaotically distributed HII  regions,  the 
NE arm seen by Thornley (1996) is  composed  of  two  mutually  crossing 
segments  having  different  pitch  angles  (Fig.~\ref{arms}).   Aligned 
H$\alpha$ structures are found only close  to  the  inner  part  of  one 
segment NE and east of  the  centre.  No  aligned  H$\alpha$  structures 
accompany the dust lanes and total power ridge on  the  western  and  SW 
disk side. 

Generally NGC~5055 seems  to  show  a  better  degree  of  alignment  of 
H$\alpha$ and  optical  features,  in  agreement  with  weak  but  still 
detectable density-wave signs described by Thornley \& Mundy (1997)  and 
by Kuno et al. (1997). The weak infrared arms discussed by  Thornley  \& 
Mundy (1997) correspond to two aligned chains of HII regions seen in our 
H$\alpha$ map. Using both the H$\alpha$ and  optical  images  we  traced 
them out to a further distance than Thornley \&  Mundy  (1997).  In  the 
enhanced optical image we identified also several spiral-like dust lanes 
(Fig.~\ref{arms}). 

To  check  for  possible  associations  between   the   magnetic   field 
orientations and those of star-forming filaments and dust lanes we again 
used  the  beam-smoothed  models  of  polarized  emission  obtained   by 
integrating the U and Q Stokes parameters due  to  an  assumed  magnetic 
field structure. The regular fields were assumed to be locally  parallel 
to  rudimentary  spiral-like  features  shown  in  Fig.~\ref{arms}   and 
distributed in a way yielding the  observed  distribution  of  polarized 
intensity. 

In NGC~3521 the mean modeled pitch angle is in  both  lobes  smaller  by 
about $3\sigma$ errors than  the  observed  one  (Fig.~\ref{lgar}).  The 
observed values of $\psi$ higher than the modeled ones occur  in  almost 
all individual map points which have the  observed  polarized  intensity 
above the noise level, the differences  reach  locally  $30\degr$  ($\ge 
6\sigma$). In the southern and SW disk of NGC~5055  (azimuths  $30\degr$ 
--  $150\degr$)  no  systematic  difference  between  mean  modeled  and 
observed magnetic pitch angles has been found. However, in the  NE  disk 
(the azimuthal angle range $180\degr$ -- $360\degr$)  the  model  yields 
small pitch angles in agreement with a preponderance of nearly azimuthal 
optical structures, while the observed mean  value  of  $\psi$  is  much 
higher, similar to that in the SW lobe  (Fig.~\ref{lgar}),  as  expected 
for an axisymmetric turbulent dynamo. We also stress that NGC~3521 while 
possessing much weaker signs of organized spiral structure does not show 
smaller mean pitch angles of polarization B-vectors  than  NGC~5055,  as 
one could expect if the radial magnetic  field  were  primarily  due  to 
local radial flows and/or compressions in arm-like spiral structures. In 
the face of the above, the radial magnetic fields and large pitch angles 
are unlikely to be caused  by  gas  flows  and  compressions  in  spiral 
features and need to be explained by a turbulent dynamo process. 

Though the dynamo action generates primarily  radial  fields,  too  high 
magnetic pitch angles can  make  a  problem  for  the  classical  dynamo 
concept. The analysis of mathematical properties of  equations  (1)  and 
(2) by Ruzmaikin et al. (1988) implies that for differentially  rotating 
galaxies (to which both discussed objects belong, Burbidge et al.  1960, 
1964): 

$${{B_r\over B_{\phi}}=\left({\beta^2\over{\alpha_0\Omega_0
h_0^3}}\right)^{1/2}} \eqno{(3)}$$ 

\noindent where $\alpha_0$ and $\Omega_0$  are  the  typical  values  of 
$\alpha$ and $\Omega$ in the disk, while $h_0$ is the  scale  height  of 
ionized gas. These authors define the dynamo number  $D$  measuring  the 
dynamo efficiency as $D = \alpha_0\Omega_0h^3_0/\beta^2$. As  the  ratio 
$B_r/B_{\phi}$ defines the magnetic pitch angle $\psi$ eq.  (3)  implies 
that: 

$${\tan \psi = |D|^{-1/2}}\propto 1/(\alpha_0\Omega_0)^{1/2}
\eqno{(4)}$$

\noindent Thus, the stronger  the  dynamo  action  the  smaller  is  the 
magnetic pitch angle. This is largely due to the fact that  the  kinetic 
helicity $\alpha$ driving the dynamo depends primarily on the turbulence 
vorticity  $<rot(\vec{v})>$  which  is   driven   by   Coriolis   forces 
proportional to the angular disk speed $\Omega$, hence  $<rot(\vec{v})>$ 
and $\alpha$ are proportional to $\Omega$,  as  well.  A  strong  dynamo 
action requires a high degree of turbulence  helicity,  and  thus  rapid 
rotation. As normal galaxies mostly rotate differentially, this in  turn 
leads to a faster conversion of the  radial  field  component  into  the 
azimuthal one (eq. (2)) thereby decreasing the magnetic pitch angle.  By 
all the above {\it large magnetic pitch angles} need a {\it weak  dynamo 
action} with a slow rotation and/or a small $\alpha$ yielding together a 
small absolute value of the dynamo number $|D|$ (eq. (4)). 

The dynamo number $D$ cannot be arbitrarily decreased to  produce  large 
magnetic pitch angles. Solutions of eq. (1) and (2) are  sought  in  the 
form  $B_{r,\phi}(t)~=~B_{r,\phi}^0e^{-\gamma~t}$.  The  magnetic   field 
generation requires its growth with time and hence a positive  value  of 
the growth rate $\gamma$. This parameter is an  increasing  function  of 
the absolute value of dynamo number $D$ (Ruzmaikin  et  al.  1988).  For 
$|D|$ smaller than some critical value $|D_{cr}|$ $\gamma$  is  negative 
and the magnetic field decays. An efficient  field  amplification  needs 
$|D|>>|D_{cr}|$.  The  latter  quantity   depends   on   many   detailed 
assumptions concerning e.g. the distribution of $\alpha$ with the height 
above the disk etc. but for realistic galaxy models $|D_{cr}|$ cannot be 
lower than $\simeq $6 -- 10. For this  reason  classical  dynamo  models 
have problems with generating spiral-like  magnetic  fields  with  pitch 
angles in excess of $17\degr$ -- $20\degr$. The above formalism  applied 
to mean magnetic pitch angles of  observed  galaxies  implies  the  mean 
dynamo numbers $|D| \approx 2.8$ in NGC~3521 and $|D|  \approx  4.5$  in 
NGC~5055, considerably below the required 6  --  10  for  normal  spiral 
galaxies (Ruzmaikin et al. 1988). Taking into account all  uncertainties 
and approximations, the classical dynamo process in the studied galaxies 
would at best work at its critical threshold.  With  $\gamma$  of  about 
zero or just above it the production  of  galaxy-scale  magnetic  fields 
would be extremely slow and inefficient, if possible at all. 

More recent, refined numerical dynamo models, involving  a  feedback  of 
magnetic fields upon  turbulent  motions  (e.g.  Elstner  et  al.  1996) 
encounter a similar pitch angle problem. The above authors  found  that, 
in the case of an axisymmetric dynamo, large pitch angles require  large 
turbulence correlation times (see  Elstner  et  al.  1996  for  detailed 
definitions), which in turn gives rise to oscillatory, dipole solutions, 
very unlikely to exist in real galactic disks. Rohde \&  Elstner  (1998) 
attained a steady dynamo solution with pitch angles  reaching  $40\degr$ 
in case of turbulence modulation by star formation enhancement in spiral 
arms. However, the applicability of this model  to  flocculent  galaxies 
having little concentration of star-forming processes in spiral arms (as 
shown by our H$\alpha$ images) is highly disputable.  The  very  recent, 
non-standard models of dynamo driven by magnetic buoyancy (Moss  et  al. 
1999) are more promising. The quoted authors obtained  typical  magnetic 
pitch angles of order $30\degr$, they also obtained reasonable solutions 
with $B_r$ comparable to $B_{\phi}$. Our  observations  of  quite  large 
pitch angles of B-vectors apparently support these new dynamo concepts. 

Another problem with applying the dynamo concept to flocculent  galaxies 
is the magnetic pitch angle asymmetry in NGC~3521.  In  normal  galaxies 
the dynamo process generates  an  axisymmetric  magnetic  field  with  a 
similar mean pitch angle in both polarized lobes  (Fig.~\ref{modpi})  as 
indeed observed in NGC~5055. In contrast,  in  NGC~3521  both  polarized 
lobes differ in the magnetic pitch angle $\psi$ by more  than  $3\sigma$ 
r.m.s. error, which cannot be explained by the axisymmetric  dynamo.  We 
note however,  that  smaller  values  of  $\psi$  in  the  SW  disk  are 
associated with smaller optical pitch angles and that this  region  also 
contains a total power ridge not caused by local star-forming  processes 
(Sect. 3). Moreover, the analysis of the red optical image by Dettmar \& 
Skiff  (1993)  suggests  that  this  region  might  recently  have  been 
perturbed  by  a  ``soft  merging''  event  or   compressions   by   the 
intergalactic gas in which case the total power ridge might result  from 
the magnetic field  squeezing.  As  the  pitch  angle  decrease  on  the 
compressed side has been observed in wind-swept galaxies NGC~4254 (Soida 
et al. 1996) and NGC~2276 (Hummel  \&  Beck  1995),  the  same  sort  of 
interactions may explain the pitch angle asymmetry in NGC~3521.

\section{Summary and conclusions}

In order to check whether  galaxies  lacking  organized  optical  spiral 
patterns still possess spiral magnetic fields, we mapped two  flocculent 
galaxies showing only weak signs of global spiral  structures  in  total 
power and polarization  at  10.55~GHz.  A  single-dish  instrument,  the 
100\,m Effelsberg telescope, was used to obtain maximum  sensitivity  to 
weak extended emission. The most important results are as follows: 

\begin{itemize}

\item[-] Both galaxies have a mean degree of magnetic  field  regularity 
      not  significantly  lower  than  that  of  a   small   sample   of 
      grand-design spirals with similar  star-forming  properties.  Some 
      differences in the polarized emission from the central regions  of 
      observed flocculent spirals may be explained by differences of the 
      star formation distribution in their inner disks 

\item[-] Despite a lack of obvious density wave action  both  flocculent 
      objects  show  spiral-like  regular   magnetic   fields   with   a 
      significant radial component as expected for  the  dynamo  action. 
      The mean magnetic pitch angles in both galaxies  are  similar  and 
      amount to some 25$\degr$ -- 30$\degr$, being  somewhat  larger  in 
      NGC~3521. Beam-smoothed polarization models were applied to  prove 
      that such large pitch angles  cannot  constitute  an  artifact  of 
      limited resolution and projection effects 

\item[-] The magnetic pitch angles are only loosely connected  to  those 
      of  rudimentary  optical  structures  and  are  likely  caused  by 
      turbulent dynamo action. However, such  large  pitch  angles  pose 
      some  problems  to  classical  dynamo  theory,  preferring   newly 
      developed non-standard dynamo concepts 

\end{itemize}

In this work we demonstrate for the first time that flocculent  galaxies 
are able to develop regular spiral magnetic fields similar  in  strength 
and structure to those in grand-design spirals,  thus  {\it  large-scale 
density-wave flows are not needed  to  produce  global  spiral  magnetic 
fields with a  substantial  radial  component}.  We  believe  that  this 
provides arguments in support  of  the  existence  of  turbulent  dynamo 
action, free from effects of density wave flows, making it difficult  to 
single-out pure dynamo-type magnetic fields in nearby spirals.  To  make 
the dynamo possible an efficient turbulent  diffusion  of  the  magnetic 
field is required rather than the  often  assumed  nearly  perfect  flux 
freezing. However, details of the magnetic field properties  may  depend 
on the distribution of star-forming processes. We think that our present 
results make flocculent galaxies worth observing at  higher  resolution, 
to clarify the details of the magnetic field properties. 

\begin{acknowledgements}

The Authors, (J.K., M.S. and M.U.) are  indebted  to  Professor  Richard 
Wielebinski from the Max-Planck-Institut f\"ur  Radioastronomie  (MPIfR) 
in Bonn for the invitations to stay at this Institute where  substantial 
parts of this work were done. A large part of the work has been done  in 
the  framework  of  the  exchange  program  between   the   Jagiellonian 
University and Ruhr-Universit\"at Bochum. We are  particularly  grateful 
to Dr Elly M. Berkhuijsen from the MPIfR for  critical  reading  of  the 
manuscripts and her many valuable comments. We are indebted to Professor 
Anvar Shukurov from the University of Newcastle, UK for his comments  on 
the dynamo process and to Dr E. Hummel  for  providing  his  unpublished 
high-resolution map of confusing sources in NGC~3521 as well  as  to  Dr 
Jonathan Braine  from  the  University  of  Bordeaux  for  his  work  on 
improving our manuscript. We are also grateful  to  numerous  colleagues 
from the MPIfR, Astronomisches Institut  der  Ruhr-Universit\"at  Bochum 
and from the Astronomical Observatory of the Jagiellonian University  in 
Krak\'ow for their comments. This work was  supported  by  a  grant  no. 
962/P03/97/12 from the Polish Research Committee (KBN). Large  parts  of 
computations were made using the HP715 workstation at  the  Astronomical 
Observatory in  Krak\'ow,  partly  sponsored  by  the  ESO  C\&EE  grant 
A-01-116 and on the Convex-SPP machine at the Academic  Computer  Centre 
``Cyfronet''  in   Krak\'ow   (grant   no.   KBN/C3840/UJ/011/1996   and 
KBN/SPP/UJ/011/1996). 

\end{acknowledgements}


\begin{thebibliography}{}

\bibitem[1977]{baars} Baars  J.W.M.,  Genzel  R.,  Pauliny-Toth  I.I.K., 
      Witzel A., 1977, A\&A 61, 99 
\bibitem[1996]{beck96} Beck R., Brandenburg A., Moss  D.,  Shukurov  A., 
      Sokoloff D., 1996, Ann. Rev. A\&A 34, 155 
\bibitem[1999]{beck99} Beck R., Ehle M.,  Shoutenkov  V.,  Shukurov  A., 
      Sokoloff D., 1999, Nat 397, 324 
\bibitem[1964]{burb1} Burbidge E.M., Burbidge G.R., Crampin D.J.,  Rubin 
      V.C., Prendergast K.H., 1964, ApJ 139, 1058 
\bibitem[1960]{burb2} Burbidge E.M., Burbidge  G.R.,  Prendergast  K.H., 
      1960, ApJ 131, 282 
\bibitem[1987]{condon} Condon J.J., 1987, ApJS 65, 485
\bibitem[1993]{dettmar} Dettmar R-J, Skiff B.A., 1993, in The  Evolution 
      of Galaxies and Their Environment, NASA Ames Research  Center,  p. 
      251 
\bibitem[1996]{elst96} Elstner D., Ruediger G., Schultz M.,  1996,  A\&A 
      306, 740 
\bibitem[1988]{emerson88} Emerson D.T., Gr\"ave R., 1988, A\&A 190, 353 
\bibitem[1979]{emerson79} Emerson D.T., Klein U., Haslam  C.G.T.,  1979, 
      A\&A 76, 92 
\bibitem[1974]{haslam} Haslam C.G.T., 1974, A\&AS 15, 333
\bibitem[1991]{humm1} Hummel E., Beck R., Dahlem M., 1991, A\&A 248, 23 
\bibitem[1995]{hummel} Hummel E., Beck R., 1995, A\&A 303, 691 
\bibitem[1997]{kuno} Kuno N., Tosaki T., Nakai N., Nishiyama  K.,  1997, 
      PASJ 49, 275 
\bibitem[1996]{lisenfeld} Lisenfeld U., V\"olk H.J., Xu C.,  1996,  A\&A 
      306, 667 
\bibitem[1986]{morsi} Morsi H.W., Reich W., 1986, A\&A 163, 313 
\bibitem[1998]{moss} Moss D., 1998, MNRAS 297, 860
\bibitem[1999]{moss2} Moss D., Shukurov A., Sokoloff D., 1999, A\&A 343, 
      120 
\bibitem[1995]{niklas} Niklas S., Klein U., Wielebinski R.,  1995,  A\&A 
      293, 56 
\bibitem[1997]{otmian} Otmianowska-Mazur K., von Linden  S.,  Lesch  H., 
      Skupniewicz G., 1997, A\&A 323, 56 
\bibitem[1998]{rohde981} Rohde R., Elstner D., 1998, A\&A 333, 27 
\bibitem[1998]{rohde982} Rohde R., Elstner D., R\"udiger G., 1998,  A\&A 
      329, 911 
\bibitem[1988]{ruzmaikin} Ruzmaikin A.A., Shukurov A.M., Sokoloff  D.D., 
      1988, Magnetic Fields of Galaxies. Astrophys.  and  Space  Science 
      Library, Vol. 133, Kluwer Academic Publishers 
\bibitem[1961]{sandage61}  Sandage  A.,  1961,  The  Hubble   Atlas   of 
      Galaxies, Carnegie Inst. of Washington, Washington D.C. 
\bibitem[1994]{sandage94} Sandage A., Bedke J., 1994, The Carnegie Atlas 
      of Galaxies, Carnegie  Ins.  of  Washington  with  The  Flintridge 
      Foundation, Washington D.C. 
\bibitem[1993]{schimdt} Schmidt A., Wongsowijoto A., Lochner O., et~al., 
      1993, MPIfR Technical Report No. 73, MPIfR, Bonn 
\bibitem[1996]{soida96} Soida M., Urbanik M., Beck R.  1996,  A\&A  312, 
      409 
\bibitem[1999]{soida99} Soida M., Urbanik M., Beck R.,  Wielebinski  R., 
      1999 A\&A 345, 461 
\bibitem[1980]{tabara} Tabara H., Inoue M., 1980, A\&AS 39, 379 
\bibitem[1996]{thornley96} Thornley M.D., 1996 ApJ 469, 45
\bibitem[1997]{thornley97} Thornley M.D., Mundy L.G., 1997, ApJ 484, 202 
\bibitem[1988]{tully}  Tully  R.B.,  1988,  Nearby   Galaxies   Catalog, 
      Cambridge Univ. Press 
\bibitem[1989]{urbanik89} Urbanik M., Beck R.,  Klein  U.,  Gr\"ave  R., 
      1989, Ap\&SS 156, 195 
\bibitem[1997]{urbanik97} Urbanik M., Elstner D., Beck  R.,  1997,  A\&A 
      326, 465 

\end{thebibliography}
\end{document}